\documentclass[twocolumn,amsmath,amssymb,prl,10pt,nofootinbib,superscriptaddress]{revtex4}
\usepackage{ graphicx, float,amsmath, amsmath, amssymb}
\usepackage[export]{adjustbox}

\def\be{\begin{equation}}
\def\ee{\end{equation}}
\def\bea{\begin{eqnarray}}
\def\eea{\end{eqnarray}}
\def\bse{\begin{subequations}}
\def\ese{\end{subequations}}

\usepackage[breaklinks, colorlinks, citecolor=blue]{hyperref}

\usepackage[labelsep=period]{caption}
\usepackage[normalem]{ulem}
\usepackage{soul}
\usepackage{minitoc}

\usepackage{bm}

\begin{document}
\title{The holographic space-time and black hole remnants as dark matter}

\author{Aur\'elien Barrau}%
\affiliation{%
Laboratoire de Physique Subatomique et de Cosmologie, Univ. Grenoble-Alpes, CNRS/IN2P3\\
53, avenue des Martyrs, 38026 Grenoble cedex, France
}







\date{\today}
\begin{abstract} 
The holographic space-time approach to inflation provides a well defined and self-contained framework to study the early universe. Based on a quasi-local quantum gravity theory generalizing string theory beyond AdS backgrounds, it addresses fundamental questions like the arrow of time and the entropy of the initial cosmological state. It was recently argued that it also provides a naturel explanation for dark matter in the form of primordial black holes. The orignal idea however suffers from troubles that can be cured by considering instead Planck relics. This possibility is investigated and paths for observational confirmation are pointed out.
\end{abstract}
\maketitle

\section{Introduction}

The holographic space-time model, mostly developed by Banks and Fischler, provides a well defined mathematical framework compatible with both general relativity and quantum field theory. It is based on unitarity, causality, and holography, describing physics in the causal diamond approach. The relevant Hilbert space has a dimension equal to the (maximal) area on the boundary of the diamond. The basic ideas are exposed in \cite{Banks:2008ep,Banks:2010tj,Banks:2010zm,Banks:2011av,Banks:2012vha,Banks:2012ic,Banks:2012nn,Banks:2013qpa,Banks:2013fr,Banks:2015iya,Banks:2015xua,Banks:2018ypk,Banks:2020dus,Banks:2020zcr}, together with the main cosmological consequences.

On the other hand, dark matter is one of the main known problems of physics. Many astrophysical searches are being carried out (see, {\it e.g.}, \cite{Censier:2011wd,Gascon:2015caa,Mayet:2016zxu,Cirelli:2012tf,Conrad:2014tla,Gaskins:2016cha,deDiosZornoza:2021rgw} for reviews). The possible production of dark matter particles by accelerators (see, {\it e.g.}, \cite{Kahlhoefer:2017dnp,Felcini:2018osp} for reviews) is also actively considered. Theoretical ideas are countless (see, {\it e.g.}, \cite{Plehn:2017fdg,Oks:2021hef,Kisslinger:2019ysx} for reviews) and new ones appear continuously. At this stage, it fair to conclude that no consensus has emerged. Although the observational evidences for dark matter are very convincing (without fully ruling out models of modified gravity \cite{Acedo:2017dvl,Katsuragawa:2017whg,Milgrom:2019cle}), the explicit search for candidates has, so far, failed. Black holes might be among the rare dark matter candidates that are known to exist (see \cite{LIGOScientific:2016aoc,EventHorizonTelescope:2019dse,EventHorizonTelescope:2019ths} for gravitational wave and interferometric observations which add up to historical arguments based on the motion of stars, accretion disks, and jets) and therefore require a minimal amount of ``new physics".

Recently, it was suggested that primordial black holes, expected in the holographic space-time model, could account for dark matter \cite{Banks:2020dgx,Banks:2021lai}. The idea that small black holes formed in the early universe could be the main component of dark matter is not new (see, {\it e.g.}, \cite{Khlopov:1999rs,Khlopov:2008qy} for wide introductions and \cite{Villanueva-Domingo:2021spv,Carr:2020xqk,Carr:2021bzv} for recent reviews) but, as it will be explained later, it is especially ``natural" in this framework. I shall however underline two major problems in this approach, related with the required non-linear merging rate and with entropy (or cosmic-rays) overproduction. Both those drawbacks can be cured at the same time, if the focus is instead moved to black holes relics. This also makes an interesting connexion with other approaches to the very same idea \cite{Barrau:2019cuo}. Interestingly, this hypothesis might be experimentally tested in the future. 

\section{The model ans its weaknesses}

The fundamental principle of the holographic space-time is that the maximal space-like area (with a factor 1/4) of the null-surface boundary of a causal diamond is equal to the logarithm of the dimension of the Hilbert space which describes the quantum information within the diamond \cite{Banks:2018ypk}. The concept of causal diamond refers to the region to which a device on a time-like geodesics can send signals and receive a response within a given proper time interval. This hypothesis relies heavily on Jacobson's derivation of general relativity from a thermodynamical perspective \cite{Jacobson:1995ab}. Degrees of freedom localized in the bulk of the causal diamond are specific (and constrained) states of the holographic variables on the boundary \cite{Banks:2020dus}. The approach is consistent with general relativity and quantum mechanics and basically free of pathologies \cite{Banks:2020zcr}. It also gives a plausible explanation to the arrow of time \cite{Banks:2012vha} and somehow adresses Leibniz's question by pointing out that ``nothing"  is the state of maximal entropy of the Universe, as all degrees of freedom then live on the cosmological horizon -- with a dynamics scrambling information at the maximal rate compatible with causality \cite{Banks:2018ypk}. The full approach is strongly linked \cite{Banks:2020tox} with the covariant entropy principle \cite{Fischler:1998st,Bousso:1999xy,Bousso:1999cb} which generalizes usual calculations of the entropy of black holes \cite{Bekenstein:1973ur} and of de Sitter spaces \cite{Gibbons:1977mu}.

As far as cosmology is concerned, the holographic space-time model leads to inflation \cite{Banks:2015xua}, but with much less freedom than in the usual approach which relies on quite a lot of free parameters  \cite{Liddle:1999mq}. Remarkably, although the core of the theory is a singularity-free quantum cosmological model, most of its consequences can be understood as coarse grained features of the underlying model \cite{Banks:2020dgx}. The key-point is that, after inflation, the Universe evolves into a dilute gas of small black holes, with sizes comparable to the horizon size at reheating. This is not an {\it ad hoc} assumption but a mere prediction in this framework \cite{Banks:2021lai}. Most of those black holes decay and produce the usual hot Big Bang. Arguments were also given that the baryon asymmetry could naturally be accounted for in this context \cite{Banks:2021lai}. The cosmological microwave background (CMB) fluctuations are recovered and, in principle, predictions could be made for the scalar-tensor ratio. Some black holes remain and form dark matter. In addition, the cosmological constant problem can be addressed through a natural relation between asymptotically large proper time and asymptotically large area \cite{Banks:2018jqo}. The $\Lambda-$term then becomes a boundary condition.

This alternative cosmological model obviously raises many questions and remains, to say the least, puzzling. At this stage, it is   quantitatively too vague and its assumptions are too radical to fully convince. Much remains to be investigated in details before it can compete with the standard model. However, its main features and robust conceptual foundations seem to justify an interest in its viability. It is worth being considered, especially when taking into account the large amount of fruitful predictions derived from very elegant and quite simple hypotheses. 

To solve the dark matter problem (and also for its self-consistency at least up to the equilibrium time), the model requires that the very small and short lived black holes formed just after inflation merge so that much more massive black holes appear, behaving as cold dark matter. They would induce the radiation-to-matter transition at cosmic time $t_{eq}$ and then solve the known dark matter issues.\\

This appealing scenario however suffers from two major drawbacks (one of them being mentioned by the authors themselves). As the post-inflation universe is, in this model, matter dominated, fluctuations are expected to grow and become highly non-linear. The associated dynamics should lead to black hole collisions and mergers \cite{Banks:2020dgx}. This is mandatory to produce black holes massive enough to survive until the equilibrium time, and even until the contemporary epoch (dark matter is still in our neighborhood). No calculation or simulation does however support this hope. The whole picture can even be considered as deeply unlikely. If black holes are, for example, formed with an initial mass $M_i\sim 10^5M_{Pl}$, at least $10^{15}$ mergers are required to produce a single black hole not yet fully evaporated. Worst: the process should be extraordinary fast as the initial lifetime of black holes in this example is $\tau\sim 3\times 10^{-27}$ s. Even if chain reactions help (mergers of mergers make the process easier), the picture requiers extreme and unexpected processes.

In addition, the very stringent constraints on the allowed abundances of primordial black holes should be taken into account \cite{Carr:2009jm}. Defining $\beta(M_i)$ as the fraction of the energy of the Universe in primordial black holes, tight upper limits on $\beta'$ (which is related to $\beta$ by factors of order one depending on the precise physics of the gravitational collapse and on the number of effectively massless degrees of freedom) were derived. Based on a detailed calculation of the nucleosynthesis dynamics, very low bounds on $\beta'$, of the order of $10^{-24}$, can for exemple be obtained for $M_i$ in the range $10^{10}-10^{14}$ g, using $^6$Li/$^7$Li and $^3$He/D abundances \cite{Carr:2009jm}. Below those masses, the entropy density still constrains $\beta'$. As soon as initial masse approaches $10^{15}$ g, that is the one corresponding to an evaporation time comparable to the Hubble time, cosmic-rays impose $\beta'<10^{-(26-27)}$. Spectral distorsions to the CMB close the remaining gaps. In its original formulation \cite{Banks:2020dgx}, the model assumes that black holes are formed with masses around 10 g. However, the mergers will inevitably produce larger black holes that will be submitted to the previous constraints. As the (unavoidable for the models) large mergers have to be very rare (otherwise the black holes would be over-abundant), black holes around $10^{10}-10^{15}$ g should be enormously more numerous than around $10^{20}$ g (where there exists a window for dark matter). Hence the conflict with the available limits. 

The model therefore nearly faces a no-go theorem. Small black holes are formed in this context for a good reason. The apparent horizon expands in a constrained state that is the origin of all localized excitations in the Universe. In global coordinates, this corresponds to a slow roll inflation period leading to a dilute gas of primordial black holes (with masses fixed by the inflationary scale) \cite{Banks:2021lai}. However, for those black holes to account for dark matter (in addition to their possible explanation of the baryon asymmetry), one needs a large instantaneous merging rate while evading the stringent upper limits available on a huge range of masses and still leaving enough light black holes to produce the hot Big Bang. Because radiation energy density dilutes as $a^{-4}$ while black holes dilute as $a^{-3}$, there will be a substantial relative enhancement ($\sim 3\times 10^{27}T_{RH}$ where $T_{RH}$ is in Planck units) of the latter up to $t_{eq}$. This, however, does not remove the constraint that the number of required mergers to produce surviving black holes is extremely high. This is were the bounds on $\beta'$ makes the scenario highly unprobable as, if black holes as massive as $10^{20}$ g (that is $10^{25} M_{Pl}$) are produced, lighter ones will be way more abundant (they have to be for consistency) and won't escape the stringent limits.

\section{A solution}

It is however possible to overcome the previously mentioned weaknesses by focusing on black hole remnants. The Hawking mechanism  \cite{Hawking:1974sw}, and the associated black hole temperature $T_H=1/(8\pi M)$ (in Planck units), is consensual. It is understood from quite a lot of different perspectives (see, {\it e.g.}, \cite{Lambert:2013uaa} for a simple introduction) and might even have already been observed in analog systems \cite{Steinhauer:2015saa}.

What happens when the semiclassical calculation breaks down -- that is in the last stages of he evaporation -- is much less clear. Numerous arguments were given in favor of the existence of stable relics (see \cite{Zeldovich:1983cr,Aharonov:1987tp,Banks:1992ba,Banks:1992is,Barrow:1992hq,Bowick:1988xh,Casher:1991en,Coleman:1991jf,Lee:1991qs,Gibbons:1987ps,Torii:1993vm,Callan:1988hs,Myers:1988ze,Whitt:1988ax,Alexeyev:2002tg,Hossenfelder:2003dy,Maziashvili:2005pp,Xiang:2006mg,Nikolic:2006nd,Banerjee:2010sd,Ali:2012hp,Rubiera-Garcia:2013gnv,Lobo:2013adx,Dirkes:2013dca,Ali:2014xqa,Chen:2014jwq,Wen:2015kwa,Mehdipour:2016vxh,Li:2017kze,EslamPanah:2018rob,Ong:2018syk,Chamseddine:2019pux,Kuntz:2019gka,Khan:2020tdk} to mention only a few references), including by one of the ``holographic space-time" founder (although in a different context). Impressively, the arguments are extremely diversified and many different ways of considering the situation lead to the conclusion that black holes should form remnants. One of the most convincing points, relying only on known physics, was made by Giddings in \cite{Giddings:1992hh} (in the information paradox framework \cite{Mathur:2012np}): causality, locality and energy conservation imply that the time scale for the final disappearance of a black hole is larger than the age of the Universe. The status of black hole remnants in holography is however still unclear \cite{Scardigli:2010gm,Berkowitz:2016muc}.

As far as dark matter in concerned, such relics would cure the main problems of the model. The idea that dark matter could be made of black hole remnants with masses $M_{rel}\sim M_{Pl}$ was first suggested in \cite{MacGibbon:1987my}. The kind a primordial fluctuations that was assumed (basically a blue primordial power spectrum) and the associated mass spectrum are however not anymore compatible with data. It was revived in \cite{Saini:2017tsz} and in \cite{Conley:2006jg,Nakama:2018lwy} (see also references therein) for extra-dimensional models and in \cite{Dalianis:2019asr} for runaway-quintessence postinflationary scenarios, among quite a lot of other proposals (see, {\it e.g.}, \cite{Barrau:2003xp,Chen:2004ft}). In \cite{Barrau:2019cuo}, black hole remnants were considered as a dark matter candidate trying to rely (nearly) only on known physics. The holographic space-time hypothesis provides exactly the ``natural" creation mechanism that was missing. It is fair to mention that the production of larger-than-expected quantities of primordial black holes during ``standard" inflation is also being seriously considered (see \cite{Dalianis:2019asr,Lehmann:2019zgt,Byrnes:2021jka,Ozsoy:2021pws,Heydari:2021qsr,Tomberg:2021xxv,Ahmed:2021ucx,Gangopadhyay:2021kmf,Flores:2021jas,Spanos:2021hpk,Teimoori:2021thk,Rigopoulos:2021nhv,Zheng:2021vda,Solbi:2021rse,Bastero-Gil:2021fac,Solbi:2021wbo,Ng:2021hll,Gao:2020tsa,Ashoorioon:2020hln,Zhou:2020kkf,Vennin:2020kng,Ragavendra:2020sop,Ozsoy:2020kat,Conzinu:2020cke,Palma:2020ejf,Ballesteros:2020qam,Lin:2020goi,Arya:2020xeo,Ashoorioon:2019xqc,Arya:2019wck,Fu:2019ttf,Bhaumik:2019tvl,Dimopoulos:2019wew,Martin:2019nuw} for recent developments) but still relies on quite speculative assumptions -- as obviously does the holographic space-time model, but with the aim of building fully new paradigm. On the specific topic of black hole production, the holographic predictions are clear \cite{Banks:2015iya,Banks:2020dgx,Banks:2021lai}.

Photons emitted sufficiently early by decaying black holes can indeed ``produce" the hot big bang, as advocated by the holographic space-time model (and initially suggested in \cite{Zel}). The requirement that they do not exceed the observed photon-to-baryon ratio reads
\begin{equation} 
\beta'<10^9\left( \frac{M_i}{M_{Pl}} \right)^{-1},
\end{equation}
where the precise definition is $\beta'(M)=\sqrt{\gamma}(g_*/106.75)^{-1/4}$, $\gamma$ being the ratio of the black hole mass to the ``standard” particle horizon mass and $g_*$ being the number of relativistic degrees of freedom normalized at $10^{-5}$ s. This indeed means that over 9 orders of magnitude, between the Planck mass and $10^4$ g, primordial black holes can produce all of the CMB. The relation between the initial black hole mass $M_i$ and the temperature of the Universe $T$ can be straightforwardly calculated as
\begin{equation} 
M_i\sim1.8\times 10^{28}\gamma\left( \frac{T}{1~\textrm{GeV}} \right) ^{-2}~\textrm{g},
\end{equation}
where, depending on the details of the collapse, $\gamma\in [10^{-4},1]$ \cite{Carr:1975qj,Hawke:2002rf}. The mass loss rate is given by the integration of the instantaneous Hawking spectrum
\begin{equation}
\frac{dM}{dt}=-\sum_i\int\frac{\Gamma_i}{2\pi}\left( e^{\frac{Q}{T_{BH}}} -(-1)^{2s_i}) \right)^{-1}QdQ,
\end{equation}
where $s_i$ is the spin of the $i-$species and $\Gamma_i$ is the associated greybody factor (directly related to the absorption cross section). Taking into account the standard model degree of freedom leads to a lifetime:

\begin{equation}
\tau\sim 4\times 10^{-28}\left( \frac{M_i}{1~\textrm{g}} \right)^3~\textrm{s}.
\end{equation}
This differs by more than one order of magnitude from the rough estimates given in \cite{Banks:2021lai,Banks:2020dgx}. Adding new particles, {\it e.g.} twice more in a supersymmetric setting, would not drastically change the picture. This shows that as long as the inflationary scale is reasonably high, the model is self-consistant and can indeed account for the hot big bang by black hole decay. Non-linear mergers are not required and the stringent limite on $\beta(M_i)$ for masses leading to a large energy release during nucleosynthesis are evaded. The remnants are enough to account for dark matter. For a reheating temperature at the GUT scale, only $10^{-24}$ of the full energy density of the Universe in the form of black hole remnants (just after inflation) would be enough to nearly close the Universe at the equilibrium time and explain the totality of dark matter. We shall come back to this point in more details as it turns out to be quite subtle in this case.

\section{Observations and fine-tuning}

Black hole remnants with masses around the Planck mass are a good dark matter candidate that can be considered as natural in the holographic space-time model. This raises at least two questions. The first one is related with a possible experimental test of this hypothesis. Finding observational evidences is very challenging. A black hole remnant would typically weight the same as a grain of dust and the associated mean numerical density would then be extraordinary small, around $10^{-18}$ relic par cubic meter, or one relic per volume element of a million times the one of planet Earth. Even if the cross section  hopefully does not vanish for the interaction with fermionic fields in the low-energy limit \cite{MacGibbon:1990zk}, the area involved (or order $10^{-66}$ cm$^2$) makes  direct detection apparently hopeless.

An exciting possibility was pointed out in  \cite{Lehmann:2019zgt}, relying on a residual electric charge for the remnants. Although quite interesting and motivated by the random emission of charged particles by black holes associated with a sharp cutoff, the idea is highly speculative.  We focus here on another possibility associated with the coalescences of relics that might have taken place during the history of the Universe \cite{Barrau:2019cuo}. The non-linear processes mentioned in \cite{Banks:2020dgx,Banks:2021lai} are not relevant in this case.
Obviously, associated gravitational waves would have a negligible amplitude and a too high frequency for any reasonable detector. When the merging occurs, the resulting black hole however acquires a mass twice higher than the minimal one. It is natural to expect that this black hole will relax to the relic state by emitting one (or a few) Planck quanta.

The merging rate can be estimated following \cite{Nakamura:1997sm, Sasaki:2016jop}. The probability of occurence during time $t$ is given by
\begin{equation}
dP=
    \frac{3}{58} \bigg[ {\left( \frac{t}{T} \right)}^{3/37}-{\left( \frac{t}{T} \right)}^{3/8} \bigg] \frac{dt}{t},
\end{equation}
with 
\begin{equation}
T \equiv  \frac{3}{170} {\left( \frac{M_{rel}}{\rho_{rel}(z_{\rm eq})} \right)}^{4/3} {(GM_{rel})}^{-3}.
\end{equation}

The rate can easily be estimated to be
\begin{equation}
N_{mer}\sim
\frac{3H_0^2}{8\pi G} \frac{\Omega_{rel}}{M_{rel}} 
\frac{dP}{dt}\bigg|_{t_0}.
\end{equation}

Importantly, at so high energies, the Universe is basically transparent (which is not the case, for example, in the TeV or PeV ranges). The associated signal can therefore be integrated without imposing a cutoff at moderate distances. We have also taken into account that emitted photons are only a small fraction of the produced particles -- gravity is democratic and the weight is only determined by the relative number of internal degrees of freedom. For working cosmic-ray experiment, the resulting flux is too small to be detected. However, it might be observed by future giant instruments. For Euso \cite{Inoue:2009zz} (or any similar telescope looking at the atmosphere of the Earth from the space station to measure extremely high energy comic-rays), nearly an event per year is expected. If one considers planets as huge cosmic-ray detectors \cite{Rimmer:2014cia}, a dozen events per year should be measured. This is challenging but not impossible.\\

The second important question in this scenario is about fine-tuning. As noticed previously, the relative weight of any matter component in the early universe will be amplified relatively to the surrounding radiation up to the equilibrium time. In the situation considered here, this effect could actually appear as too efficient and become a problem for the model. Quite impressively, this is however not the case. Let us be more specific. The Zel'dovich bound requires the initial mass of black holes to be smaller than $10^9M_{Pl}$ to account for all the CMB. It is however easy to convince oneself that, in the considered context, the upper end of this mass interval is favored. Too small black holes would indeed leave remnants only a few orders of magnitude lighter than the original mass and would therefore contribute way to much to the energy budget of the early universe (leading to an unreasonably small equilibrium time). On the other hand, black holes in the range $10^6M_{Pl}<M_i<10^9M_{Pl}$ leave appropriate remnants. The initial relative energy of the Universe in the form of relics (that is $10^{-9}-10^{-6}$) seems still way too high but several effects add up to cure this. First, and most importantly, the evaporation cannot anymore be assumed to be sudden. Not only does the Hawking time scales as $M_i^3$ but one should expect the evaporation to  ``slow down" a few orders of magnitude above $M_{Pl}$ (this has been, in particular, studied in string gravity \cite{Alexeyev:2002tg}). Second, it is probable that inflation does not end suddenly \cite{Garcia:2020eof}, delaying the formation of some black holes. Third, the critical phenomena presumably associated with the collapse \cite{Hawke:2002rf} can decrease $\gamma$ up to $10^{-4}$ which subsequently increases the cosmic time associated with a given initial black hole mass. It should also be noticed that the evaporation dynamics is such that the first stages last much longer than the last ones, the Universe therefore remaining matter dominated for most of the lifetime of the black holes. All in one, depending on several parameters still poorly determined, the upper end of the mass interval compatible with producing all the CMB with decaying black holes could generate remnants with the appropriate density to explain dark matter. Especially when taking into account that, if a gas of black holes is the favored quantum state in the considered framework, this does not mean that only black holes are formed at the end of inflation.

Very nicely, the initial masses predicted by the holographic model, independently of the above arguments, is $10^6M_{Pl}<M_i<10^9M_{Pl}$. This is simply derived by fitting the CMB \cite{Banks:2020dgx,Banks:2021lai} with a slow-roll parameter $\epsilon\in [10^{-4},10^{-1}]$.  This also matches the so-called ``triple coincidence" \cite{Alexander:2007gj}.

The model requires fine-tuning in the sense that if the black holes had been formed at a different mass -- corresponding to a different inflationary scale -- the equilibrium time would have been different. It should be stressed that this is not a problem. There are 2 very different fine-tuning issues. One merely says that no attractor is involved and states that ``if things had been otherwise in the past, the present would be different". This is true in most physical situation and this is {\it not} problematic. The other one basically says that we are in a very specific situation -- from a bayesian point of view -- and that this objectively exceptional state wouldn't have been accessed for the vast majority of reasonable initial conditions. This is problematic. Otherwise stated: when playing a huge roulette with, say, $10^{50}$ sections and spinning the ball there is nothing wrong with ending up with a result having a tiny probability of $10^{-50}$ unless this corresponds to the only {\it a priori} specific and relevant result. In this latter case, one should indeed suspect that something hidden has driven the ball. 

In cosmology, $\Omega=1$ is a very specific situation -- euclidean geometry. But inflation imposes a (nearly) vanishing curvature: as $(\Omega^{-1}-1)=-\frac{3k}{8\pi \rho a^2}$ with $\rho$ being constant as the scale factor increases by many e-folds, $\Omega$ is driven to 1. There is clearly no magics here \cite{Barrau:2019cuo} as $\Omega$ includes a normalization to the critical density which depends on the Hubble constant. This means that a different occurence of the model would have changed the equilibrium time and the amount of dark matter -- that are fully contingent quantities -- but {\it not} the ``objectively" special features of our cosmological situation. In this sense, there is no fine-tuning problem.

\section{Conclusion}

The holographic spacetime model intents to provide a quasi-local theory of quantum gravity generalizing string theory beyond asymptotically flat and anti-de Sitter backgrounds \cite{Banks:2013qpa}. It leads to a non-singular cosmological model from the Big Bang to the post-inflationary era. Recently, it was claimed to also account for dark matter. It seems that this eventuality might be more convincing if focusing on black hole remnants.

Obviously, a lot remains to be investigated and this note just aimed at drawing a first possible line of development, among many others. It would, in particular, be very fruitful to compare the entropic considerations of the holographic space-time with recent ideas put forward by Rovelli on a different but very natural ground \cite{Rovelli:2018vvy}.

This brief article just aimed at pointing out a possibly promising direction of research. But, even for this specific idea, much remains to be done to ensure consistency. Most orders of magnitude were here set by implicitly assuming the usual scenario and the associated values of the relevant parameters. For most of it, the new paradigm remains to be built. 

\bibliography{refs}

 \end{document}